\documentclass[prl,twocolumn,superscriptaddress,floatfix]{revtex4}
\usepackage{graphicx,amsfonts,amssymb,amsmath, hyperref}

\newif\ifhyper
\hypertrue
\ifhyper
\hypersetup{
   citecolor = {green},
   colorlinks = {true}, % false
   urlcolor = {blue} % magenta
}
\fi

\newcommand{\beq}{\begin{equation}}
\newcommand{\eeq}{\end{equation}}
\newcommand{\beqa}{\begin{eqnarray}}
\newcommand{\eeqa}{\end{eqnarray}}
\newcommand{\ket} [1] {\vert #1 \rangle}
\newcommand{\bra} [1] {\langle #1 \vert}

\def\bra#1{\langle#1\vert}
\def\ket#1{\vert#1\rangle}

\def\ipr#1#2{\langle#1\vert#2\rangle}

\def\Longarrow{\protect\@lra}
\def\@lra{\relbar\joinrel\relbar\joinrel\relbar\joinrel%
          \relbar\joinrel\rightarrow}

\begin{document}

\title{Topological Geometric Entanglement}

\author{Rom\'an Or\'us}
\affiliation{School of Mathematics and Physics, The University of Queensland,
QLD 4072, Australia}
\affiliation{Max-Planck-Institut f\"ur Quantenoptik, Hans-Kopfermann-Str. 1, 85748
Garching, Germany}

\author{Tzu-Chieh Wei}
\affiliation{Department of Physics and Astronomy, University of
British Columbia, Vancouver, BC V6T 1Z1, Canada} \affiliation{C. N.
Yang Institute for Theoretical Physics, State University of New York
at Stony Brook, NY 11794-3840, USA}

\begin{abstract}

Here we show the connection between topological order and the geometric entanglement, as measured by the logarithm of the overlap between a given state and its closest product state of blocks, for the topological universality class of the toric code model. As happens for the entanglement entropy, we find that for large block sizes the geometric entanglement is, up to possible subleading corrections, the sum of two contributions: a non-universal bulk contribution  obeying a boundary law times the number of blocks, and a universal contribution quantifying the underlying pattern of long-range entanglement of a topologically-ordered state. 

\end{abstract}
%\pacs{05.30.-d, 02.70.-c, 03.67.Mn, 05.50.+q}

\maketitle

{\it Introduction.-} Topological order (TO) \cite{to} is an example of new physics beyond Landau's symmetry-breaking paradigm of phase transitions. Systems exhibiting this new kind of order are linked to concepts of the deepest physical interest, e.g. quasiparticle anyonic statistics and topological quantum computation \cite{topoq}. Importantly, TO finds a realization in terms of Topological Quantum Field Theories \cite{tqft}, which are the low-energy limit of quantum lattice models such as the toric code and quantum double models \cite{toric}, as well as string-net models \cite{sn}.  

A remarkable property about TO is that it influences the long-range entanglement in the wave function of the system. For instance, as proven for systems in two spatial dimensions (2D) \cite{Hamma, entr}, the entanglement entropy $S$ of a block of boundary size $L \gg 1$ obeys the law $S = S_0 - S_{\gamma} + O(1/L)$,
where $S_0 \propto L$ is a non-universal term (the so-called ``boundary law"), and $S_{\gamma}$ is a universal long-distance contribution: the topological entanglement entropy. $S_{\gamma}$ is non-zero for systems with TO, e.g. $S_{\gamma} = 1$ for systems in the topological universality class of the toric code. More recently, similar universal contributions have also been found for other bipartite entanglement measures \cite{mut, reny}. 

The purpose of this letter is to investigate, in the context of systems with TO, a global measure of entanglement which captures truly non-bipartite correlations in the system: \emph{the geometric entanglement (GE)} \cite{ge}, which we call $E_G$. For simplicity, we focus on systems in the topological universality class of the toric code \cite{toric}. Our results indicate that, generically, the geometric entanglement of \emph{blocks} of boundary size $L \gg 1$ obeys the law
\beq
E_G = E_0 - E_{\gamma} + O(1/L),
\label{toge}
\eeq
where $E_0 \propto n_b L$ is a non-universal term that obeys some boundary law scaling multiplied by the number of blocks $n_b$, and $E_{\gamma}$ is the \emph{topological geometric entanglement}.  Moreover, we find that $E_{\gamma}= S_{\gamma}$, with $S_{\gamma}$ the topological entanglement entropy. These results are also the first explicit example of boundary laws for the GE in ground states of 2D quantum many-body systems. 

{\it The toric code topological universality class.-}
Let us start by reviewing some basics on the toric code model \cite{toric}. This is the renormalization group (RG) fixed point of the topological universality class of a $\mathbb{Z}_2$ lattice gauge theory, and is equivalent under local transformations to the Levin-Wen string model on a honeycomb lattice \cite{sn} (see the Appendix).

We consider a square lattice $\Sigma$ on a torus. Other Riemann surfaces of genus $\mathfrak{g}$ could also be considered easily without changing our conclusions. Non-bipartite lattices (e.g. honeycomb) could also be considered without changing the long-distance properties. There are spin-1/2 (qubits) degrees of freedom attached to each link in $\Sigma$. The model is described in terms of stars and plaquettes. A star ``$+$" is a set of links sharing a common vertex. A plaquette ``$\square$" is an elementary face on the lattice $\Sigma$. For any star $+$ and plaquette $\square$, we consider the star operators $A_+$ and plaquette operators $B_{\square}$ defined as  $A_+ \equiv \prod_{j \in +} \sigma_x^{[j]}$ and $B_{\square} \equiv \prod_{j \in \square} \sigma_z^{[j]}$, where $\sigma_{\alpha}^{[j]}$ is the $\alpha$-th Pauli matrix at link $j$ of the lattice. Let us call respectively $n_+$, $n_{\square}$ and $n$ the number of stars, plaquettes and links in lattice $\Sigma$. Star and plaquette operators satisfy the global constraint
\beq
\prod_+ A_+ = \prod_{\square} B_{\square} = \mathbb{I}. 
\label{con}
\eeq
Therefore, there are $n_+-1$ independent star operators and $n_{\square}-1$ independent plaquette operators. 

With the definitions above, the Hamiltonian of the model reads $H = -\sum_+ A_+ -\sum_{\square} B_{\square}$. This Hamiltonian can be diagonalized exactly as explained in Ref.~\cite{toric}. The ground level of $H$ is $4$-fold degenerate ($4\mathfrak{g}$-fold for a Riemann surface of genus $\mathfrak{g}$). This degeneracy depends on the underlying topology of $\Sigma$, and is a signature of topological order. Moreover, the ground level is a stabilized space of $G$, the group of all the possible products of independent star operators, of size $|G| = 2^{(n_+ - 1)}$. 

In order to build  a basis for the ground level subspace, let us consider a closed curve $\gamma$ running on the links of the dual lattice $\Sigma^*$. We define its associated loop operator $W_x[\gamma] = \prod_{j \in \gamma} \sigma_x^{[j]}$, where $j \in \gamma$ are the links in $\Sigma$ crossed by the curve $\gamma$ connecting the centres of the plaquettes. It is not difficult to see that the group $G$ can also be understood as the group generated by all contractible loop operators \footnote{Notice that some sets of non-contractible loops are also elements of $G$, such as two parallel non-contractible ones. This is because they can be built from products of contractible loops.}. Let also $\gamma_1$ and $\gamma_2$ be the two non-contractible loops on a torus, and define the associated string operators $w_{1,2} \equiv W_x[\gamma_{1,2}]$. We call $\ket{0}$ and $\ket{1}$ the eigenstates of $\sigma_z$ respectively  with $+1$ and $-1$ eigenvalue. The ground level subspace then reads $\mathcal{L} = {\rm span} \{ \ket{i,j}, ~ i,j=0,1 \}$, where 
\beq
\ket{i,j} = |G|^{-1/2} \sum_{g \in G} g ~ w_1^i ~ w_2^j ~ \ket{0}^{\otimes n}.
\label{gs}
\eeq
It is easy to check that the four vectors $\ket{i,j}$ are orthonormal and stabilized by $G$, that is, $g \ket{i,j} = \ket{i,j} ~ \forall g \in G$ and $\forall i,j$. These four states form a possible basis of the ground level subspace of the model on a torus. 

Excited states can be constructed by locally applying Pauli operators $\sigma_x, \sigma_y, \sigma_z$ on the ground states $\ket{i,j}$. Pauli operators $\sigma_z$ create pairs of deconfined charge-anticharge quasiparticle excitations,  $\sigma_x$ create pairs of deconfined flux-antiflux quasiparticles, and $\sigma_y$ creates a flux-antiflux and a charge-anticharge pairs. These operators can be applied to several sites of the lattice, thus creating a quasiparticle pattern that defines the excited state. The excited states of the model are then labeled by a set of quantum numbers, $\ket{\phi,c, i,j}$, where $\phi$ and $c$ are patterns denoting the position of flux-type and charge-type excitations respectively, and $i,j$ label the ground state $\ket{i,j}$ that was excited. In this paper we will focus on the entanglement properties of states $\ket{\phi,c, i,j}$.

\begin{figure}
\includegraphics[width=0.43\textwidth]{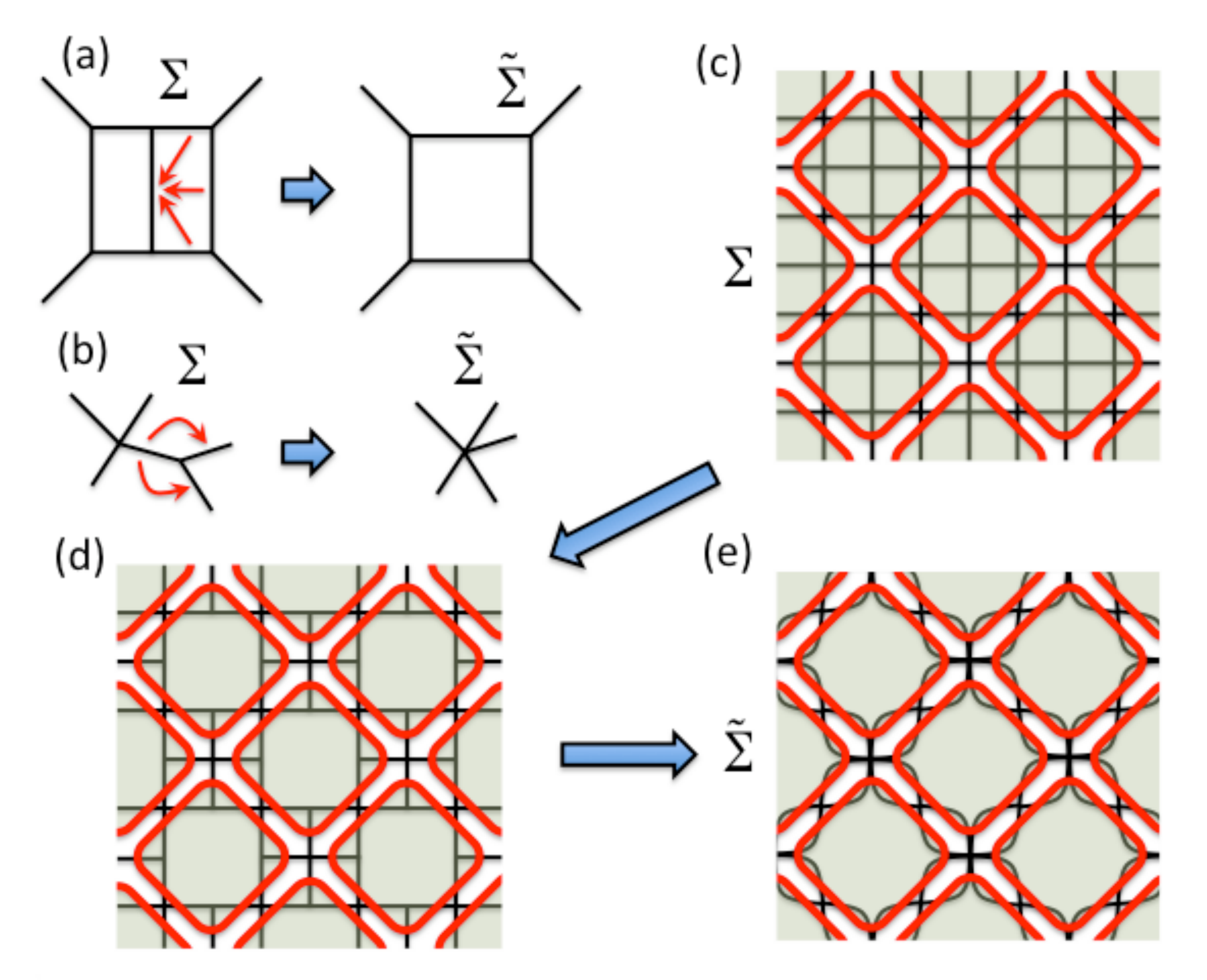}
\caption{(color online) (a,b) CNOT operations that disentangle qubits from the system, thus removing their links from the lattice. In the diagram, the arrows go from controlling to target qubits; (c) Example of a partition into blocks of a given boundary $L$, e.g. $L = 12$ here (but all our derivations work for arbitrary $L$). Qubits crossed by the boundary of a block are regarded as being inside the block; (d,e) Doing CNOT operations locally inside of each block we can remove all the stars inside of all the blocks in two steps: first, we apply CNOTs as in (a) to get (d) and, second, we apply CNOTs as in (b) to get (e). Notice that in (e) each star is made of either 4 or 8 qubits.  The procedure works similar for other blockings.}
\label{fig:diag}
\end{figure}

A key property of the toric code model is that its ground state $\ket{0,0} \equiv \ket{i=0,j=0}$ can be created by a quantum circuit that applies a sequence of Controlled-NOT (CNOT) unitary operations over an initial separable state of all the qubits \cite{cnots}. This means that it is actually possible to \emph{disentangle} qubits from the ground state of the system by reversing the action of these CNOTs. This was the key observation that allowed to build an exact MERA representation of the ground states $\ket{i,j}$ of the model \cite{tcmera}. The two fundamental disentangling movements are represented in Fig~\ref{fig:diag}(a,b), and leave the overall quantum state as a product state of the disentangled qubits with the rest of the system. What is more, the rest of the system is left in the ground state of a toric code model on a deformed lattice $\widetilde{\Sigma}$, where $\widetilde{\Sigma}$ is obtained from $\Sigma$ by removing the links that correspond to the disentangled qubits. This property turns out to be of great importance for some of the derivations in this paper. 

{\it Geometric entanglement.-} Let us now remind the basics of the geometric entanglement. Consider an $m$-partite normalized pure state $\ket{\Psi} \in \mathcal{H} = \bigotimes_{i=1}^{m} \mathcal{H}^{[i]}$, where $\mathcal{H}^{[i]}$ is the Hilbert space of party
$i$. For instance, in a system of $n$ spins each party could be a single spin, so that $m = n$, but
could also be a set of spins, either contiguous (a \emph{block} \cite{geometric2}) or not. We wish now to determine how well state $\ket{\Psi}$ can be
approximated by an unentangled (normalized) state of the parties, 
$\ket{\Phi}\equiv\mathop{\otimes}_{i=1}^{m}|\phi^{[i]}\rangle$. The proximity of $\ket{\Psi}$ to $\ket{\Phi}$ is captured by their
overlap. The entanglement of $\ket{\Psi}$ is thus revealed by the maximal
overlap~\cite{ge}, $\Lambda_{\max}({\Psi})\equiv\max_{\Phi}|\ipr{\Phi}{\Psi}|$.
The larger $\Lambda_{\max}$ is, the less entangled is $\ket{\Psi}$. We quantify the entanglement of
$\ket{\Psi}$ via the quantity:
\begin{equation}
E_G({\Psi})\equiv-\log_2\Lambda^2_{\max}(\Psi), \label{eq:Entrelate}
\end{equation}
where we have taken the base-2 logarithm, and which gives zero for unentangled states. $E_G(\Psi)$ is called \emph{geometric entanglement} (GE). This quantity has been studied in a variety of contexts, including critical systems and quantum phase transitions \cite{geometric2, geometric3}, quantification of entanglement as a resource for quantum computation \cite{resource}, local state discrimination \cite{discrim}, and has been recently measured in NMR experiments \cite{exper}.  Also, one can choose the case of just two sets of spins. In this case the GE $E_G(\Psi)$ coincides with the so-called single-copy entanglement  between the two sets, $E_1(\Psi) = - \log_2 \nu_1(\rho)$, with $\nu_1(\rho)$ the largest eigenvalue of the reduced density matrix $\rho$ of either set \cite{sc}. 

The GE offers a lot of flexibility to study multipartite quantum correlations in spin systems. For instance, one can choose each party to be a single spin, but one can also choose blocks of increasing boundary length $L$ \cite{geometric2} (see Fig.~\ref{fig:diag}(c) for an example). Studying how the GE changes with $L$ provides information about how close the system is to a product state under coarse-graining transformations. 

{\it GE of the toric code universality class.-} We now study the GE of the topological universalty class of the toric code model. 
By definition of topological universality class, the long-distance properties of all the states in this class are equivalent to those of the RG fixed point, that is, the toric code model. Hence, if we are interested in extracting the universal topological (long-distance) contribution to the GE of the states in this class, it is sufficient to focus our analysis on the fixed point only. 

We thus consider the entanglement properties of the four ground level states $\ket{i,j}$ as well as the excited states $\ket{\phi,c,i,j}$. Our main result here is that, for these states, the GE of blocks consists of a boundary term plus a topological contribution. In order to see this we first consider a couple of useful Lemmas: 

\begin{figure}
\includegraphics[width=0.43\textwidth]{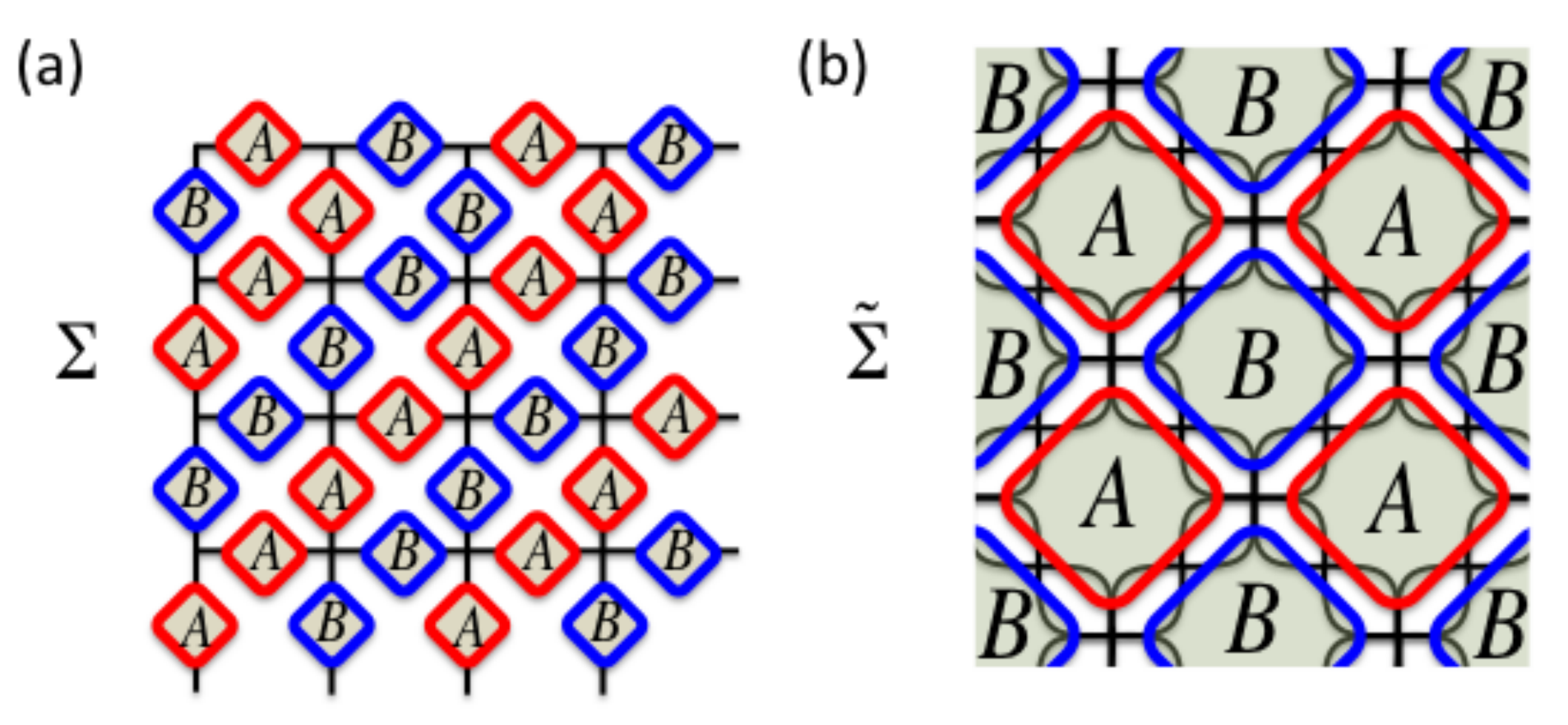}
\caption{(color online) Bipartite lattices of spins and blocks: (a) Sets of spins $A$ and $B$ for Theorem 1 and a $4 \times 4$ torus; (b) Sets of blocks $A$ and $B$ for Theorem 2.}
\label{part}
\end{figure}

{\it {\bf Lemma 1:} All the eigenstates $\ket{\phi, c, i, j}$ of the toric code Hamiltonian have the same entanglement properties}. 

{\it Proof:} let us choose a reference state from the ground level basis, e.g. $\ket{0,0}$. First of all, notice that the four basis states $\ket{i,j}$ in the ground level are related to $\ket{0,0}$ by local unitary operators,  since the string operators $w_1$ and $w_2$ are tensor products of $2 \times 2$ identity operators and Pauli-$x$ matrices. Since these operators act locally on each spin, they do not change the entanglement of the quantum state. Moreover, the excited states $\ket{\phi,c,i,j}$ are created locally by applying tensor products of $2 \times 2$ identity and Pauli-$x$ and $z$ operators to $\ket{i,j}$. The states resulting from these operations have then the entanglement properties of state $\ket{i,j}$, which in turn are the same as those of the reference state $\ket{0,0}$. Thus, all states $\ket{\phi, c, i, j}$ have the same entanglement properties. $\square$

{\it {\bf Lemma 2:} Consider an arbitrary set of blocks of qubits in lattice $\Sigma$, where each block is regarded as an individual party. Then, the entanglement properties of the ground state $\ket{0,0}$ are the same as those of state $\ket{\widetilde{0,0}}$, the ground state of the toric code model in the deformed lattice $\widetilde{\Sigma}$ obtained after disentangling as many qubits as possible using CNOTs locally inside of each block.} 

{\it Proof:} given the ground state $\ket{0,0}$ in $\Sigma$, we start by doing CNOT disentangling operations \emph{locally inside of each block} in order to disentangle as many qubits as possible. As a result, state $\ket{0,0}$ transforms into state $\ket{0,0}_{{\rm disentangled}} \equiv \ket{e_1} \otimes \cdots \otimes \ket{e_p} \otimes \ket{\widetilde{0,0}}$, where  $\ket{e_k}$ is the quantum state for the $k$-th disentangled qubit, $k = 1,\ldots, p$ ($p$ is the number of disentangled qubits), and $\ket{\widetilde{0,0}}$ is the $i=0,j=0$ ground state of a toric code Hamiltonian in the deformed lattice $\widetilde{\Sigma}$. Since all CNOT operations are done locally inside of each block, states $\ket{0,0}$ and $\ket{0,0}_{{\rm disentangled}}$ have the same entanglement content if we regard the blocks as individual parties. What is more, the entanglement in the system is entirely due to the qubits in $\ket{\widetilde{0,0}}$, which proves the lemma. $\square$

Keeping the above two lemmas in mind, we now present the following theorem about the geometric entanglement of spins in the toric code: 

{\it {\bf Theorem 1 (for spins):} For the toric code Hamiltonian in lattice $\Sigma$, the four ground states $\ket{i,j}$ for $i,j = 0,1$ and also the excited states $\ket{\phi, c, i, j}$ have all the same GE of spins and is given by
\beq
E_G = n_+ - 1, 
\label{getc}
\eeq
where $n_+$ is the number of stars in $\Sigma$.}

{\it Proof:}  Using Lemma 1 we can restrict our attention to the GE for the state $\ket{0,0}$. In this setting, we call \emph{computational basis} the basis of the many-body Hilbert space constructed from the tensor products of the $\{ \ket{0}, \ket{1} \}$ local basis for every spin, which is an example of \emph{product basis} for the spins. 

Our proof follows from upper and lower bounding the quantity $E_G(0,0)$ in Eq.(\ref{eq:Entrelate}) for the ground state $\ket{0,0}$. First, from the expression in Eq.(\ref{gs}) for the ground states we immediately have that the absolute value of the overlap with any state of the computational basis is $|G|^{-1/2}$, and therefore $\Lambda_{\max}\ge |G|^{-1/2}$. From here, we get $E_G(0,0) \le  \log_2|G|$, 
which gives an upper bound. 

Next, to derive a lower bound 
we use the fact that if we group the $n$ spins into two sets $A$ and $B$, then $\Lambda_{\max}\le \Lambda_{\max}^{[A:B]}$,
where $[A:B]$ means that a partition of the system with respect to the two sets $A$ and $B$ is considered. Thus, we have that  $E_G(0,0)\ge E_G(0,0)^{[A:B]}$. 

The trick to find a useful lower bound is to find an appropriate choice of sets $A$ and $B$. In our case, we consider e.g. the bipartiton shown in Fig.\ref{part}(a) for even $\times$ even lattices (other cases can be considered similarly). Then, we use a Lemma by Hamma, Ionicioiu and Zanardi in Ref.\cite{Hamma},
that the reduced density matrix $\rho_A$ of $|0,0\rangle$ for subsystem $A$ satisfies $\rho_A^2= (|G_A| |G_B| / |G| )\rho_A$,
where $G_{A/B}$ is the subgroup of $G$ acting trivially on subsystem
$A/B$. Importantly, for $A$ and $B$ chosen as explained above, it happens that 
$G_A$ and $G_B$ are \emph{trivial groups consisting of only the identity element}. With this in mind, we see that the reduced density matrix $\rho_A$ has eigenvalues either zero or
$|G|^{-1}$, which is $|G|$-fold degenerate. This means that
$({\Lambda}_{\max}^{[A:B]})^2=|G|^{-1}$ and hence $E_G(0,0)\ge \log_2|G|$.

Combining the two bounds, we get $E_G(0,0)= \log_2|G|=n_+-1$, and from here Eq.(\ref{getc}) for the GE for spins follows immediately. $\square$

Importantly, the techniques used in Theorem 1 can also be used to deal with blocks of spins whenever the blocks form a bipartite lattice. In general, one may first disentangle as many qubits as possible inside the blocks. Then the remaining qubits are left in an entangled state where the GE is equal to the number of remaining independent star operators, which amounts to a boundary law term (possibly with a subleading correction depending on how the blocks are chosen) plus a topological term. As an example of this, let us present the following theorem:

{\it {\bf Theorem 2 (for blocks):} Given a partition of the lattice $\Sigma$ into $n_b$ blocks of boundary size $L$ as indicated in Fig.~\ref{fig:diag}(c)  (where $L$ is measured in number of qubits), then the four ground states $\ket{i,j}$ for $i,j = 0,1$ and also the excited states $\ket{\phi, c, i, j}$ of the toric code Hamiltonian have all the same GE of blocks and is given by 
\beq
E_G = (n_bL/4) - 1.
\label{getcL}
\eeq}

{\it Proof:} First, notice that Lemma 1 and Lemma 2 imply that we can entirely focus on the ground state $\ket{\widetilde{0,0}}$ of a toric code model in the deformed lattice $\widetilde{\Sigma}$ from Fig.~\ref{fig:diag}(e). As shown in Fig.~\ref{fig:diag}(d,e), it is always possible to remove \emph{all} the stars inside of each block  in $\Sigma$ just by doing CNOTs locally inside of each block. As a result, the stars in the deformed lattice $\widetilde{\Sigma}$ correspond to those in $\Sigma$ that lay among the blocks. It is easy to see that, for $n_b$ blocks of boundary $L$, there are $\widetilde{n}_+ = n_b L/4$ of such stars.

The rest of the proof follows as in Theorem 1, by upper and lower bounding $E_G(0,0)$ for the partition with respect to blocks. For the upper bound, we use the fact that ${\Lambda}_{\max}^{[{\rm blocks}]} \ge {\Lambda}_{\max}^{[{\rm spins}]} \ge |\widetilde{G}|^{-1/2}$, where the first inequality again used the property that if larger blocks are considered then the maximum overlap is also larger, and $\widetilde{G}$ is the corresponding group of contractible loop operators on $\widetilde{\Sigma}$. From here $E_G(0,0) \le  \log_2|\widetilde{G}|$ follows. For the lower bound, we divide the system into two sets $A$ and $B$ of blocks as indicated in Fig.\ref{part}(b). Following a similar reasoning as in Theorem 1, it is possible to see again that no element $g \in \widetilde{G}$ will act trivially on $A$ or $B$ except for the identity element. From this point, the rest of the proof is simply equivalent to the proof for Theorem 1.  $\square$

{\it Discussion.-} Let us now discuss Eqs.~(\ref{getc}) and (\ref{getcL}) in detail. First, let us remind that a variety of works have shown the existence of a "boundary law" for the entanglement entropy of many 2D systems \cite{boun}, including models with TO \cite{Hamma, entr}. It is then remarkable that, according to the first term of these equations, the entanglement per block obeys also a boundary law. To the best of our knowledge, these results are the first example of a boundary law behaviour for a multipartite (rather than bipartite) measure of entanglement in 2D. 

The second term in Eqs.~(\ref{getc}) and (\ref{getcL}) is far more intriguing and important. Its existence is caused by the global constraint from Eq.~(\ref{con}) on star operators which, in turn, allow for the topological degeneracy of the ground state. Thus, this term is \emph{of topological nature and universal}, and quantifies the pattern of long-range entanglement present in topologically ordered states. 

 In order to clarify further the meaning of the topological term, let us consider the case of just two blocks of spins. In this case, the geometric entanglement $E_G$ coincides with the single-copy entanglement $E_1 = -\log \nu_1(\rho)$, with $\nu_1(\rho)$ the largest eigenvalue of the reduced density matrix $\rho$ of either block. Since this density matrix is proportional to a projector (see Ref.~\cite{Hamma} and also Ref.~\cite{reny}), we have that $E_1 = S$, with $S$ the entanglement entropy of the bipartition. Thus $E_G = E_1 = S$. We also know that for a system with TO, the entanglement entropy satisfies $S = S_0 - S_{\gamma}$, where $S_0$ is some boundary law term, and $S_{\gamma}$ is the topological entropy. As explained in Ref.\cite{Hamma}, for the toric code model the global constraints in Eq.(\ref{con}) imply that $S_{\gamma}=1$. Thus, with the convention from Eq.(\ref{toge}), we have that the \emph{topological geometric entanglement} is given by $E_{\gamma} = S_{\gamma}$. 

\begin{figure*}
\includegraphics[width=0.7\textwidth]{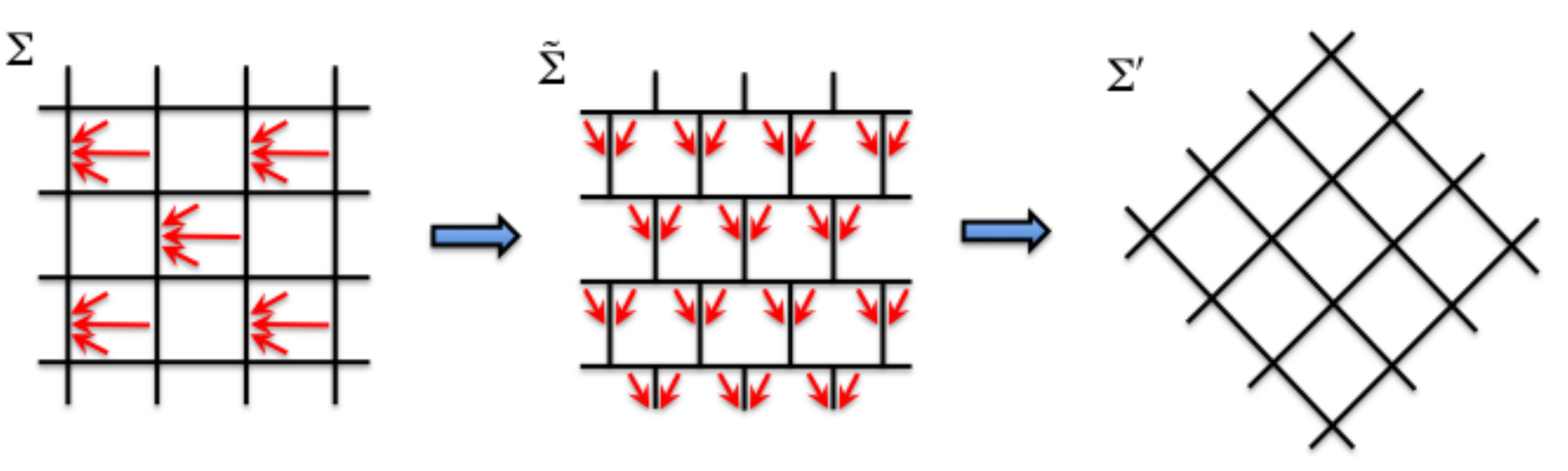}
\caption{(color online) Mapping of the toric code from a square lattice $\Sigma$ to a honeycomb lattice $\widetilde{\Sigma}$ (Levin-Wen string model), and back to a square lattice $\Sigma'$. Each red arrow represent a CNOT operation that disentangles qubits from the system, as explained in Fig.1(a,b) of the main paper. The arrows go from controlling to target qubits.}
\label{figMap}
\end{figure*}

{\it Robustness of $E_{\gamma}$ away from the fixed point.-} 
There are several ways of checking the robustness of $E_{\gamma}$ under perturbations. One possibility is to perform a numerical analysis. This is not considered here, and will be the subject of a future work \cite{fut}. Another option is a perturbation theory analysis, which we present in detail in the Appendix. Yet, a more intuitive alternative is the following argumentation based on RG fixed points:  the toric code on a square lattice is an RG fixed point and hence representative of its topological universality class. The long-distance properties of any state in this universality class do not change under local RG transformations and,  thus,  are equivalent to those of the fixed point. This, in particular, is true for the long-range pattern of entanglement, and hence for $E_{\gamma}$. Thus, any non-relevant and short-range perturbation driving the Hamiltonian away from the fixed point will produce ground states with the same $E_{\gamma}$. Nevertheless, we expect a change in the short-range pattern of entanglement, and hence in the non-universal boundary law for $E_0$. For short-range perturbations the change involves modifications of the prefactor of the boundary law as well as the possible appearance of subleading $O(1/L)$ corrections. This can be checked e.g. by perturbation theory (see the Appendix). Hence, Eq.(\ref{toge}) applies when away from the fixed point. 

{\it Conclusions.-} Here we have studied the GE of blocks in topologically-ordered states. We found that it is composed of a boundary law term plus a topological term, apart from possible subleading corrections. We focused on the topological universality class of the toric code model, which includes the simplest string-net model of Levin and Wen \cite{sn}. We believe that similar results should also apply to the A-phase of Kitaev's honeycomb model \cite{honey} for which the Toric Code is (in some limiting cases) an effective model, as well as spin-liquid states with an emergent $\mathbb{Z}_2$ gauge symmetry \cite{z2}. 

\acknowledgements
We thank M. Aguado, O. Buerschaper, W.-M. Son, H.-H. Tu and G. Vidal for illuminating discussions and insightful comments. Support from UQ, ARC and EU are acknowledged. 

\appendix

\section{Appendix}

In this Appendix we provide the following further information: first, we explain an explicit local mapping that shows the local equivalence of the toric code model on a square lattice and the Levin-Wen string model on the honeycomb lattice. Hence, bot models are valid RG fixed points representing the same topological universality class. Second, we perform a perturbation theory analysis of the robustness of the GE of the toric code model under external short-range perturbations such as magnetic fields. This analysis provides upper and lower bounds for the GE, and complements the RG argumentation on the robustness of $E_{\gamma}$ presented in the paper. 

\subsection{(i) Local mapping between the square and honeycomb lattices}

The toric code model on a honeyconb lattice is, in fact, the simplest of the string-net models of Levin and Wen \cite{sn}. By definition, this model describes the fixed point of a topological phase capturing all the long-range properties of the universality class of a $\mathbb{Z}_2$ lattice gauge theory. Here we show that, in fact, the toric code model on a square lattice is also a valid RG fixed point. A way to see this is to realize that it is actually possible to map the model on the square lattice to the honeycomb lattice, and back again to the square, by means of a sequence of disentangling (coarse-graining) CNOT transformations acting locally on the lattice. These are represented in the diagram of Fig.(\ref{figMap}). We stress, though, that it is already well known that the toric code in a square lattice is an RG fixed point, see e.g. the Entanglement Renormalization analysis from Ref.(\cite{tcmera}). 

\subsection{(ii) Perturbation theory analysis of the robustness}

Let us now add a perturbation to the toric code Hamiltonian on the square lattice, and see how the ground state changes. For simplicity, we consider the case of an infinite plane, where the ground state $\ket{0,0}$ is non-degenerate, and hence we can use non-degenerate perturbation theory. The perturbed Hamiltonian will be 

\beq
H^{\lambda} = H + \lambda V , 
\eeq
where $H$ is the toric code Hamiltonian, $V$ is the perturbation, and $\lambda \ll 1$. Non-degenerate perturbation theory says that the new ground state can be approximated as

\beq
\ket{0,0}^{\lambda} \approx \ket{0,0} + \lambda \sum_{\phi,c} \frac{\bra{\phi,c,0,0}V\ket{0,0}}{E_{0,0} - E_{\phi,c}} \ket{\phi,c,0,0} \, 
\eeq
where $E_{0,0}$ is the ground state energy and $E_{\phi,c}$ is the energy of the excited state $\ket{\phi,c,0,0}$. 

We now consider the case in which the perturbation is an homogeneous magnetic field, e.g. in the $x$ direction,  

\beq 
V = \sum_{j = 1}^n \sigma_{x}^{[j]} 
\eeq
(the case of $z$ and $y$ directions can be considered similarly). It is easy to check that in this case, the normalized perturbed ground state becomes

\beq
\ket{0,0}^{\lambda} \approx C\left( \ket{0,0}  - \frac{\lambda}{\Delta} \sum_{j = 1}^n \sigma_{x}^{[j]} \ket{0,0} \right) \ ,
\eeq
with $\Delta$ the energy gap to create a pair of flux and anti-flux quasiparticles, and $C = (1+n \lambda^2/\Delta^2)^{-1/2}$ a normalization constant. 

Our aim now is to estimate the maximum overlap of the pervious state with a product state of blocks of boundary size $L$. This can be done as follows: first, and as in the unperturbed case, we apply CNOT operations locally inside of the blocks so that qubits are disentangled in the unperturbed ground state. By doing this, we can focus on the entanglement of the state

\begin{widetext}

\beq
\ket{0,0}^{\lambda}_{{\rm disentangled}} \approx C\left( \ket{\widetilde{0,0}}  - \frac{\lambda}{\Delta} \sum_{j = 1}^n \sigma_{x}^{[j]} \ket{\widetilde{0,0}} \right) \otimes \ket{e_1} \otimes \cdots \otimes \ket{e_p} \ ,
\eeq
where $\ket{e_k}$ is the quantum state for the $k$-th disentangled qubit. The above equation is indeed equivalent to

\beq
\ket{0,0}^{\lambda}_{{\rm disentangled}} \approx C\left( \ket{\widetilde{0,0}} \otimes \ket{e_1} \otimes \cdots \otimes \ket{e_p} 
 - \frac{\lambda}{\Delta} \sum_{j = 1}^{n_b} S_{x}^{[j]} \ket{\widetilde{0,0}} \otimes \ket{e_1} \otimes \cdots \otimes \ket{e_p} - \frac{\lambda}{\Delta}  \ket{\widetilde{0,0}}  \ket{\omega_{1,\dots, p}}\right)\ ,
\eeq
\end{widetext}
where $S_{x}^{[j]}$ is the total spin in the $x$ direction for the $L$ spins in the boundary of block $j$, and 
\beq
\ket{\omega_{1,\dots, p}} = \sum_{j=1}^p  \sigma_{x}^{[j]}  \ket{e_1} \otimes \cdots \otimes \ket{e_p} \ . 
\eeq
Now we find upper and lower bounds to the maximum overlap of this state with a product state of the blocks. A lower bound can be easily obtained by the product state $\ket{0}^{\otimes (n-p)} \otimes  \ket{e_1} \otimes \cdots \otimes \ket{e_p}$. Noticing that $\ket{e_k}$ is either $\ket{0}$ or $\ket{+}$ \cite{tcmera}, and that the $\ket{+}$ contributions come only from qubits close to the boundary of the block, we have that 

\beq
C|\widetilde{G}|^{-1/2} \left(1 -  \frac{\omega n_b L \lambda}{\Delta} \right) \le \Lambda_{{\rm max}}^{{\rm [blocks]}} \  ,  
\eeq
for some positive $\omega = O(1)$ constant. The following upper bound can also be found easily: 
\beq
 \Lambda_{{\rm max}}^{{\rm [blocks]}} \le C |\widetilde{G}|^{-1/2} \left( 1 + \frac{n_b L \lambda}{\Delta} + \frac{\lambda}{\Delta}\right) \ . 
 \eeq
Using the above bounds, one can check that for the GE we obtain, in the limit $\lambda \ll \Delta$ and $L \gg 1$, 
 \beq
\left(\frac{1}{4} + \frac{2 \omega \lambda}{\Delta}\right) n_b L- 1 \ge E_{G}^{\lambda} \ge \left(\frac{1}{4} - \frac{2 \lambda}{\Delta}\right) n_b L - 1 \ . 
 \eeq
The above equation is compatible with a leading change in the GE in the prefactor of the boundary law. Also, the fact that both bounds leave the topological component $E_{\gamma} = 1$ untouched seems to indicate that this is actually robust under the perturbation. Moreover, implementing finite-$L$ corrections to these bounds provides $O(1/L)$ corrections, and thus Eq.(1) in the main paper applies.

\end{document}